\documentclass[review, times]{elsarticle}

\usepackage{lineno,hyperref}
\usepackage[normalem]{ulem}
\modulolinenumbers[5]
\usepackage{graphicx}
\usepackage{subfigure}
\usepackage{color}
\usepackage{color}
 \usepackage{url}

\begin{document}

\begin{frontmatter}

\title{Low-pressure THGEM-based operation with Ne$+$H$_{2}$ Penning mixtures}

\author[ND]{ J. S. Randhawa}
\ead{jrandhaw@nd.edu}
\author[ND]{T. Ahn\corref{mycorrespondingauthor1}}
\cortext[mycorrespondingauthor1]{Corresponding author.}
\ead{tan.ahn@nd.edu}

\author[ND]{J. J. Kolata}
\author[ND]{P. O'Malley}
\author[ND,CO]{A. Ontiveros}

\address[ND]{Department of Physics and Astronomy, University of Notre Dame, Notre Dame, Indiana 46556-5670, USA}
\address[CO]{University of Northern Colorado, Greeley, CO 80639 }


\begin{abstract}

The operation of Thick Gas Electron Multipliers (THGEMs)  in Time Projection Chambers (TPCs) using various gas mixtures has applications in various nuclear and particle physics experiments. Of particular interest in low-energy nuclear physics is study of nuclear reactions involving Ne isotopes. These reactions can be studied using Ne-based gas mixtures at low pressures in TPCs used as active targets for radioactive beams. We report on the low-pressure operation of THGEMs in Ne + H$_{2}$ gas mixtures. We show that, since the Ne+H$_{2}$ forms a Penning pair, higher gains with THGEMs are achievable compared to pure neon gas. Moreover,  H$_{2}$ acts as a quench gas allowing for higher THGEMs voltages while producing  minimal background in reactions such as fusion compared to carbon-based neon gas mixtures. Detailed electron transport and amplification simulations have been performed and they qualitatively agree with the increased gain H$_{2}$ provides in the Ne:H$_{2}$ mixture. The higher THGEM gains that are achieved with a Ne:H$_2$ mixture will enhance the study of Ne-based reactions in active-target detectors that have high-granularity pad planes, leading to higher spatial-resolution heavy-ion track data.
 
\end{abstract}


\end{frontmatter}


\section{Introduction}
\label{sec1}
Time Projection Chambers (TPCs) used as active targets are at the forefront of detector technologies for studies in nuclear physics with radioactive ion beams \cite{Bazin,Ayyad2018}. 
The Gas Electron Multiplier (GEM)\cite{Sauli97} and THick Gas Electron Multiplier (THGEM) \cite{Breskin2009} are a frequently used and robust Micro-Pattern Gas Detector (MPGD). Though the functioning of GEMs and THGEMs in various gases have been studied in detail \cite{Breskin2009,Cortesi2007}, GEM/THGEM operation in noble gases has been of special interest. Out of various noble gases, of particular interest has been the use of pure Ne gas. High-pressure ($\sim$200 bar) Ne gas has been considered as a target medium in dark matter searches and for coherent neutrino scattering \cite{White05}, and Ne in its supercritical phase has the potential to detect low-energy solar neutrinos \cite{Galea07}. In this context, GEM based operations have been explored in high-pressure Ne gas and also at cryogenic temperatures in previous work\cite{Buz2005}.\\
\\
In contrast, for studies in low-energy nuclear physics, operating GEMs with Ne gas at low pressure is needed \cite{Avila16}. 
Ne gas as a target is of particular interest as Ne-Ne and Ne-Mg fusion cross-sections play a critical role in  pycnonuclear burning in the crust of accreting neutron stars \cite{Yakovlev10,Beard10}. In pycnonuclear burning, ions settle into a lattice structure in the crust of the accreting neutron star so the only energy available is zero-point lattice vibrations. Therefore, relevant energies for pycnonuclear burning, for which cross-section data is required, are very low. Measurements of cross sections below the Coulomb barrier are required as they determine the extrapolations of the cross sections to zero energy. Active-target time projection chambers provide a useful tool to measure cross-sections below the Coulomb barrier \cite{Kolata} . High detection efficiency ($\sim$100\%) and good vertex resolution allows for the measurement of fusion cross-sections well below the Coulomb barrier.
In previous studies, THGEMs operation in Ne+CH$_{4}$  \cite{Cortesi09} and Ne+iC$_{4}$H$_{10}$ \cite{Ne_iso} gas mixtures at atmospheric pressures have resulted in high gains ($\approx$10$^{5}$). However, these mixtures are not well suited for Ne fusion reactions in TPCs as the carbon in CH$_{4}$ and iC$_{4}$H$_{10}$ leads to background fusion reactions. Therefore, to measure Ne+Ne and Ne+Mg fusion cross-sections at low energy, use of pure Ne as a target gas in TPCs is desired as it can provide background free data. However,  as the operation of GEMs or THGEMs need very high electric fields across a small amplification region and hence the neon gas reaches the sparking limit before a sufficient gain is achieved for pressures that are needed for low-energy reaction studies. 

In this work, we focus on demonstrating the use of 
Ne:H$_{2}$ mixtures as an alternative to pure neon gas, as the H$_{2}$ in a Ne:H$_{2}$ gas mixture has a negligible effect on fusion cross-section measurements due to the large mass and charge differences between neon isotopes and protons. The protons and Ne nuclei result in very different energy loss and therefore signals, allowing them to be easily separated. The main advantage of using Ne:H$_{2}$ gas mixtures is that  Ne and H$_{2}$ constitute a Penning pair where the energy of the metastable state of Ne (i.e. Ne$^{m}$) is higher than the ionization energy of H$_{2}$. Therefore, the following reactions can significantly enhance the ionization coefficient:  
\begin{equation}
    e+\mathrm{Ne}\rightarrow e+\mathrm{Ne^{m}}
\end{equation}
\begin{equation}
\mathrm{Ne^{m}+H_{2}}\rightarrow \mathrm{Ne+H_{2}^{+}}+e.
\end{equation}
H$_{2}$ also allows for higher voltages to be reached before sparking due to its role as a quench gas. These hydrogen based gas mixtures have been studied earlier using triple-GEM detectors at cryogenic temperatures where it was shown that Ne+H$_{2}$ gas mixtures provide higher gain compared to pure Ne gas \cite{Buz2005}. Here we report on room temperature low-pressure GEM-based operation in Ne:H$_{2}$ gas mixtures. Specifically, we use a two-layer multi-layer THGEM ( or M-THGEM) \cite{Cortesi2017} for our study, which we will refer to from now on as THGEM for brevity. We measured the gain versus amplification voltage characteristics in Ne:H$_{2}$ (95:5) and Ne:H$_{2}$ (98:2) gas mixtures  at various low pressures (150-300 Torr). We compared the obtained results to the simulated THGEM response using detailed microscopic electron transport simulations. Section \ref{sec2} provides the details about the experimental setup to measure the gain in Ne:H$_{2}$ mixtures, Section \ref{sec3} show the results of current measurements, and Section \ref{sec4} provides details of the simulations and shows comparison to the data. Section \ref{sec5} summarizes the current work with an outlook for future studies and applications.

\section{Experimental set-up}

\begin{figure}
\centering
\includegraphics[width=\linewidth]{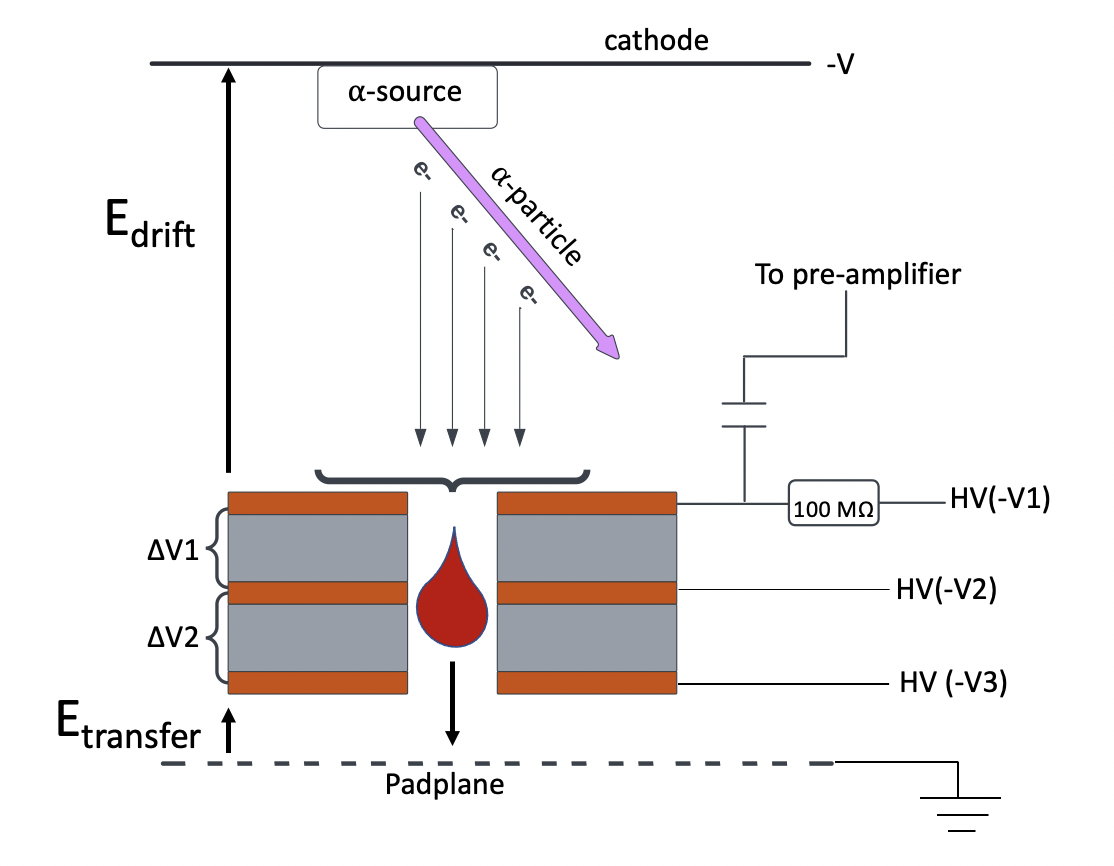}
\caption{A schematic of the experimental set-up (not to scale) is shown which includes the drift region, THGEM and transfer region. THGEM has three layers of copper conductor and two insulating layers consisting of halogen-free laminate. The $\alpha$-source  was placed 26 cm above the top of the THGEM, which was placed $\sim$2mm above the anode pad. For the gain measurement, THGEMs were operated in symmetric mode i.e. $\Delta$ V$_{1}$=$\Delta$ V$_{2}$.  }
\label{figure1}
\end{figure}

We have used the newly commissioned Notre-Dame Cube (ND-Cube), an active target TPC, to study low-pressure THGEM operation in Ne+H$_{2}$ gas mixtures. Detailed information on the ND-Cube can be found in reference \cite{Ahn2022}. A two-layer M-THGEM was used for primary electron amplification, and was placed $\approx$2 mm above the pad-plane. The THGEM was produced at CERN and has three layers of copper conductor separated by two insulating layers consisting of halogen-free laminate. The structure is illustrated in Fig.~\ref{figure1}. The primary electron amplification occurs in the regions between the conducting layers. The total thickness of the THGEM is 1.2 mm and the holes are 0.5 mm in diameter with a 0.1 mm rim and 1.0 mm pitch. The outer conductors were given a Ni/Au finish. To measure the gain as a function of amplification voltage across the THGEM layers, we used  a $^{148}$Gd   $\alpha$-source having an  activity  of  2000  Bq. We  placed  the  source 26 cm above the THGEM. Two different mixtures of Ne+H$_{2}$ gas, i.e. Ne:H$_{2}$ (95:5) and (98:2)  were used. Gain measurements were performed at 150, 200, and 300 Torr for the Ne:H$_{2}$ (95:5) mixture and at 200 Torr and 300 Torr for the Ne:H$_{2}$ (98:2) mixture. 
For the pressures used in these measurements, the $\alpha$-particles ($E= 3.2$ MeV) from $^{148}$Gd were completely stopped inside the active area. 

 The THGEM gains for the Ne:H$_2$ mixtures were measured using a capacitavely-coupled charge-sensitive preamplifier to integrate the induced current on the upper electrode of the THGEM. Simulations using Garfield++ \cite{garf} indicate that this measurement of the integrated current, which is the effective gain, is within about 10\% of the number of electron-ion pairs created in a Townsend avalanche, the true gain. The measured effective gain would be somewhat smaller for a measurement on the pad plane due to electrons collected on the THGEM's lower electrode. The larger signal resulting from the top THGEM electrode was advantageous. In addition, the top electrode signal is free from the influence of the choice of the transfer field $E_{\rm trans}$. The transfer region with $E_{\rm trans}$ is shown in Fig.~\ref{figure1}. During normal operation, each individual anode pad in the pad plane would be connected to the front-end electronics of a data acquisition system such as the GET electronics \cite{Pollaco2018}. The preamplifier pulse was shaped with a spectroscopic amplifier with a 10 $\mu$s shaping time, which sufficiently covers the induced signal time scale of 5 $\mu$s. The total charge was calibrated by using a pulser and a capacitor with a known capacitance. The gain was calculated from the initial number of primary electrons, which was derived from the known energy of the 3.2 MeV $\alpha$ particle and its calculated ionization density. The electric field in the drift region was fixed at 30 V/cm. The THGEM was operated in a symmetric mode where the voltages across the two amplification regions were kept the same, i.e., $\Delta$V$_{1}=\Delta$V$_{2}$ in Figure 1.  

\label{sec2}



\section{Results}
\label{sec3}

\begin{figure}
\centering
\includegraphics[width=\linewidth]{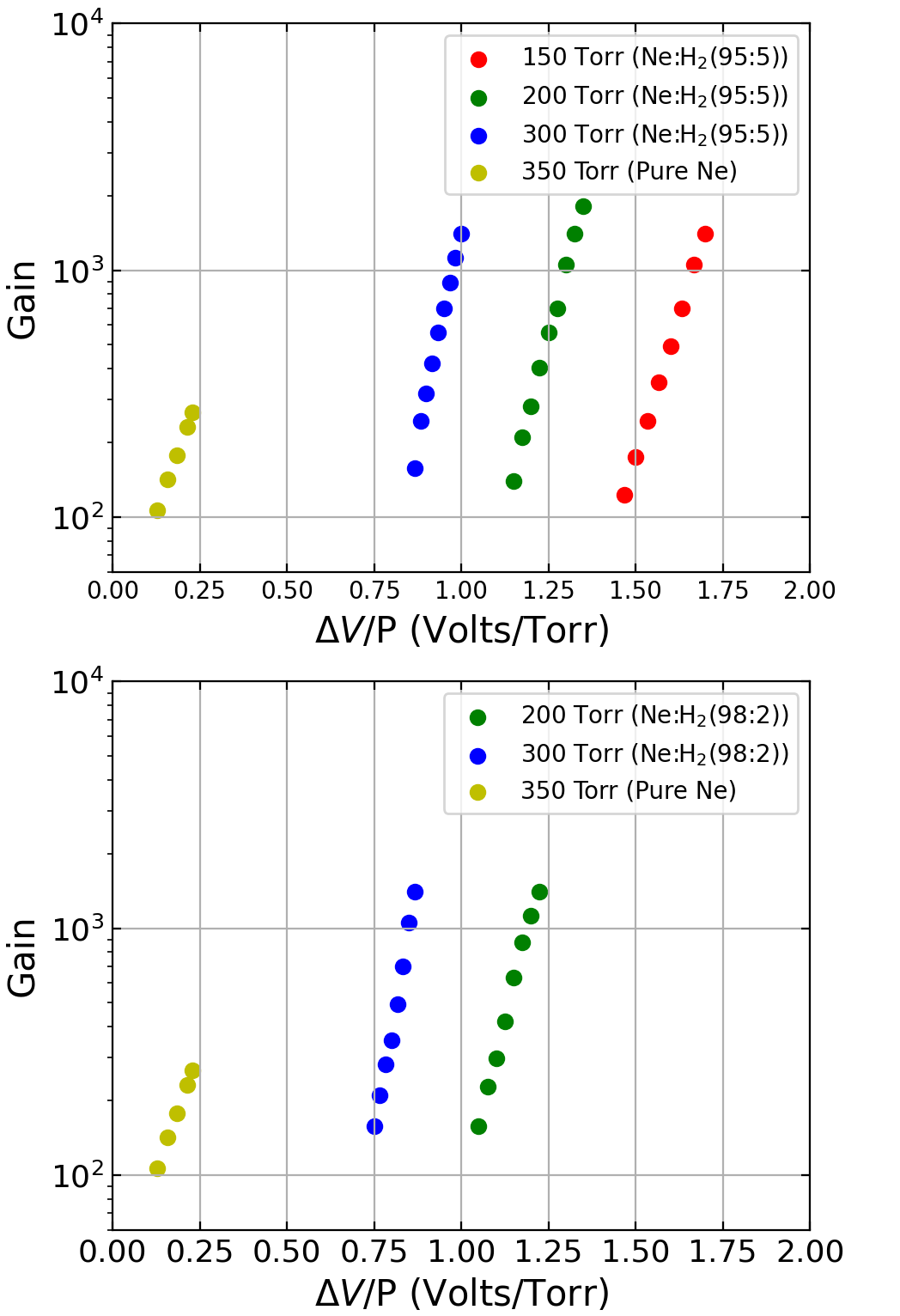}
\caption{Measured gain in two Ne mixtures: the Ne:H$_{2}$ (95:5) (upper panel) and Ne:H$_{2}$ (98:2) (lower panel) at different pressures. With Ne:H$_{2}$ (95:5), the measurements could be extended down to 150 Torr.}
\label{figure2}
\end{figure}

The measured gains  as a function of the reduced voltage ($\Delta$V/P where $\Delta V=\Delta V_{1}=\Delta V_{2}$) are shown in Fig.~\ref{figure2} for the gas mixtures Ne:H$_{2}$ (95:5) and Ne:H$_{2}$ (98:2). \color{black}For the Ne:H$_{2}$ (95:5) mixture, the gain was measured at three different pressures: 150 Torr, 200 Torr and 300 Torr. For the Ne:H$_{2}$ (98:2) mixture, gains were measured at 200 Torr and 300 Torr as the measurements at 150 Torr were limited by the onset of discharge leading to amplifcation instabilities. In both gas mixtures, gains of 10$^{2}$-10$^{3}$ were obtained for different amplification voltages, with a maximum gain of more than 10$^{3}$ for both mixtures for the pressures considered in this study. In comparison, the gain in pure Ne gas has a  maximum gain of $\approx$300 before the onset of discharge (see Fig.~\ref{figure2}).
In addition, the lowest pressures at which a significant gain could be achieved for pure Ne was 350 Torr. Lower pressures were limited by discharge.
As measurements with the Ne+H$_{2}$ (95:5) mixture were possible down to 150 Torr, this shows H$_{2}$ also serves as a quench gas allowing for higher voltages to be achieved.  
However, from this comparison it is difficult to infer if there is a gain enhancement in Ne+H$_{2}$ mixture compared to the pure Ne gas at the same pressure and electric field strength.
To address this question, we have provided a detailed comparison of Townsend coefficients in Section~\ref{subsec3_1} to infer the gain enhancement in a Ne+H$_{2}$ mixture. In Section~\ref{subsec3_2}, we show an $\alpha$-particle image as an example to demonstrate that the gains achieved in Ne+H$_{2}$ mixtures are sufficient to image charged particles with a comparable ionization density. This example can be used to benchmark future measurements such as the imaging of heavy-ion beams and fusion products including light particles.

\subsection{Reduced Townsend coefficients and comparison with pure Ne gas}
\label{subsec3_1}

\begin{figure}[h]
\centering
\includegraphics[width=\linewidth]{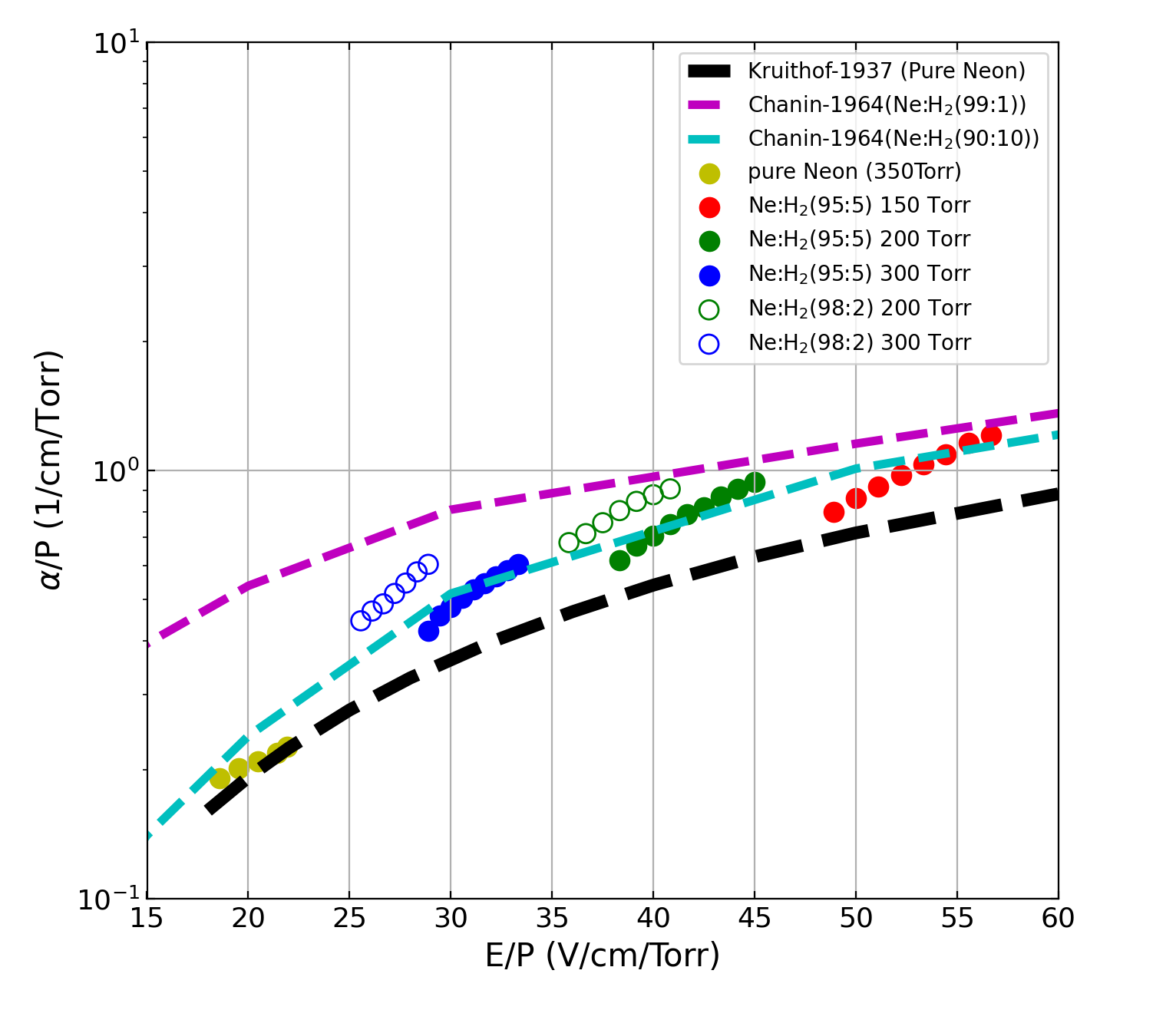}
\caption{Reduced ionization coefficients as a function of reduced field. Ionization coefficients for Ne+H$_{2}$ agrees with previous measurements and are enhanced with respect to pure neon.}
\label{figure3}
\end{figure}

 In the work of Chanin \textit{et al.}~\cite{Chanin1964}, primary ionization coefficients were measured in Ne+H$_{2}$ mixtures, with the H$_{2}$ content varying from 0.01\% to 10\%. In their work, Chanin \textit{et al.} showed that one of the most significant characteristics of these Penning mixtures is the enhancement of the ionization coefficient over that of each constituent, i.e., either pure neon or pure hydrogen. \color{black} To compare our data with previous measurements, we obtained the  ionization coefficients from the measured gain following the procedure of Ref.~\cite{Buz2005}. To obtain the reduced Townsend coefficients from the measured gain,  a parallel-plate approximation to the avalanche development inside the THGEM amplification region was applied. As we are using a dual-layer THGEM, we assumed the top and bottom electrode to be two parallel plates  with $\Delta V= \Delta V_1+ \Delta V_2$, where $\Delta V_1 =  \Delta V_2$. \color{black} Therefore, the electric field was uniform inside the amplification region.  As the gain G $= e^{\alpha d}$, where $\alpha$ is the Townsend ionization coefficient and $d$ is the total amplification region distance, the reduced  Townsend ionization coefficients can be written as
\begin{equation}
   \alpha/p = \frac{\ln G}{p d}, 
\end{equation}
 where $p$ is the pressure in Torr. Fig.~\ref{figure3} shows the reduced Townsend ionization coefficient as a function of the reduced electric field. 
 For consistency in our comparison to literature values, we normalized our measured values of the reduced Townsend coefficients to match those of Kruitof and Penning \cite{Kruithof1937} for pure Ne (Figure~\ref{figure3} dashed black line). The reduced ionization
 coefficients for our Ne:H$_{2}$ (98:2) mixture are higher than the values for the Ne:H$_{2}$ (90:10) gas mixture given by Chanin \textit{et al}.\ and is consistent with their observation of a increase in the Townsend
 coefficients for decreasing H$_{2}$ concentrations from 10\% to 1\%. The reduced Townsend coefficients for the Ne:H$_{2}$ (95:5) mixture deduced in this work (Figure~\ref{figure3} filled circles) overlaps with
 the  values for Ne:H$_{2}$ (90:10) (Figure~\ref{figure3} dashed cyan line) providing a reasonable agreement. The reduced coefficients with the Ne:H$_{2}$ (98:2) mixture in our work (Figure~\ref{figure3} open circles) are closer to the values obtained in Chanin \textit{et al}.\ for the
 Ne:H$_{2}$ (99:1) mixture (Figure~\ref{figure3} dashed magenta line). Overall, the reduced ionization coefficients obtained for both gas mixtures are higher compared to pure Ne, as expected for a Penning mixture. This enhancement in gain for a given pressure and electric field, and the suppression of discharge by H$_2$ show the advantage of a Ne:H$_2$ mixture over pure Ne at pressures around 200 Torr.\\

\subsection{Imaging with THGEMs}
\label{subsec3_2}
\begin{figure}[h]
\centering
\includegraphics[width=\linewidth]{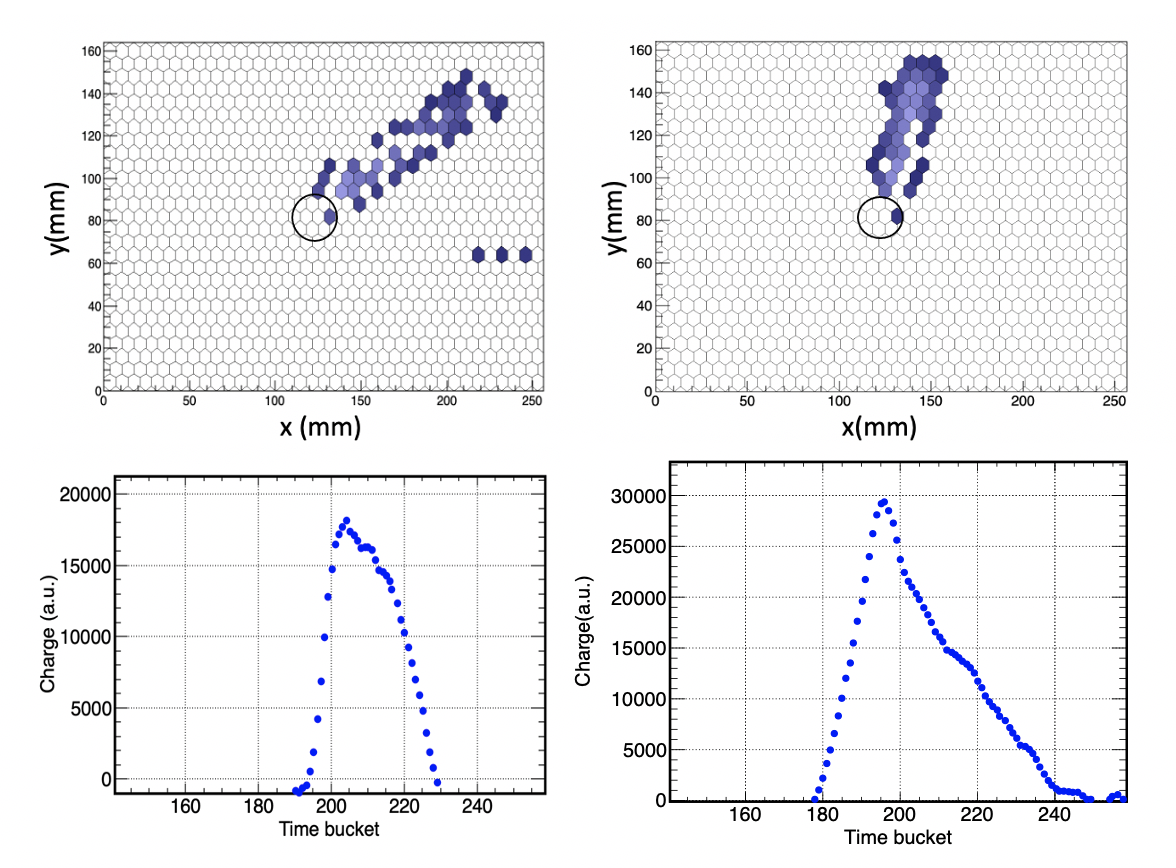}
\caption{Imaging of 3.2 MeV $\alpha$-particles using THGEMs with Ne+H$_{2}$(95:5) gas mixtures.}
\label{figure4}
\end{figure}

 THGEMs are among one of the most robust and readily used MPGDs used in imaging detectors. Therefore, to demonstrate the use of THGEMs as imaging detectors with a Ne:H$_{2}$ Penning mixture, we provide the images of $\alpha$-particles from a $^{148}$Gd source. The source was placed 26 cm above the top layer of our THGEM and the ND-Cube was filled with 200 Torr of the Ne:H$_{2}$ (95:5) mixture. The amplification voltage in each region of the THGEM was set to $\Delta V_1=\Delta V_2= 250$ V which corresponds to 2080 V/cm. The transfer voltage between the bottom layer of the THGEM and the pad plane was kept at 1000 V/cm. The imaged $\alpha$-particle and the corresponding Bragg curves are shown in Fig.~\ref{figure4}, where the X-axis given in units of time bins represents the uncalibrated position along the axis normal to the pad plane. \color{black}  As the $\alpha$-source is facing downward, the two example images shown are for $\alpha$-particles emitted at two different angles  but completely stopped inside the active area covered by pad plane. The two-dimensional projection of $\alpha$ tracks can be clearly seen in Fig.~\ref{figure4} despite the fact that electron diffusion is known to be larger in  pure Ne and Ne-based gas mixtures (e.g. Ne+CH$_{4}$) \cite{Cortesi09} compared to lighter gases. The interest in these Penning mixtures is their use as a target for fusion reactions. As demonstrated in the recent ND-Cube commissioning \cite{Ahn2022}, the ionization densities of the beam particles (heavier than helium) as well as their fusion products will be much higher compared to the ionization densities of $\alpha$-particles. For example, the average energy-loss of a 3.2 MeV $\alpha$-particle in Ne+H$_{2}$ at 200 Torr is on the order of 25 keV/mm. In comparison the average energy loss of a 40 MeV $^{24}$Mg beam is approximately 250 keV/mm, where $^{24}$Mg was chosen as a representative ion for in-beam reaction studies. As the number of primary electrons produced are proportional to the energy loss,  heavier recoils, fusion products, and beam particles can sufficiently be imaged together with lighter particles, e.g.~ $\alpha$-particles. Moreover, using this $\alpha$-particle image as an example case, one can benchmark or estimate the expected signal for future experiments with Ne+H$_{2}$ mixtures by comparing to our anode pad size of 0.4 cm$^2$.

\section{Simulations and comparison with the data}
\label{sec4}

\begin{figure}
\centering
\includegraphics[width=\linewidth]{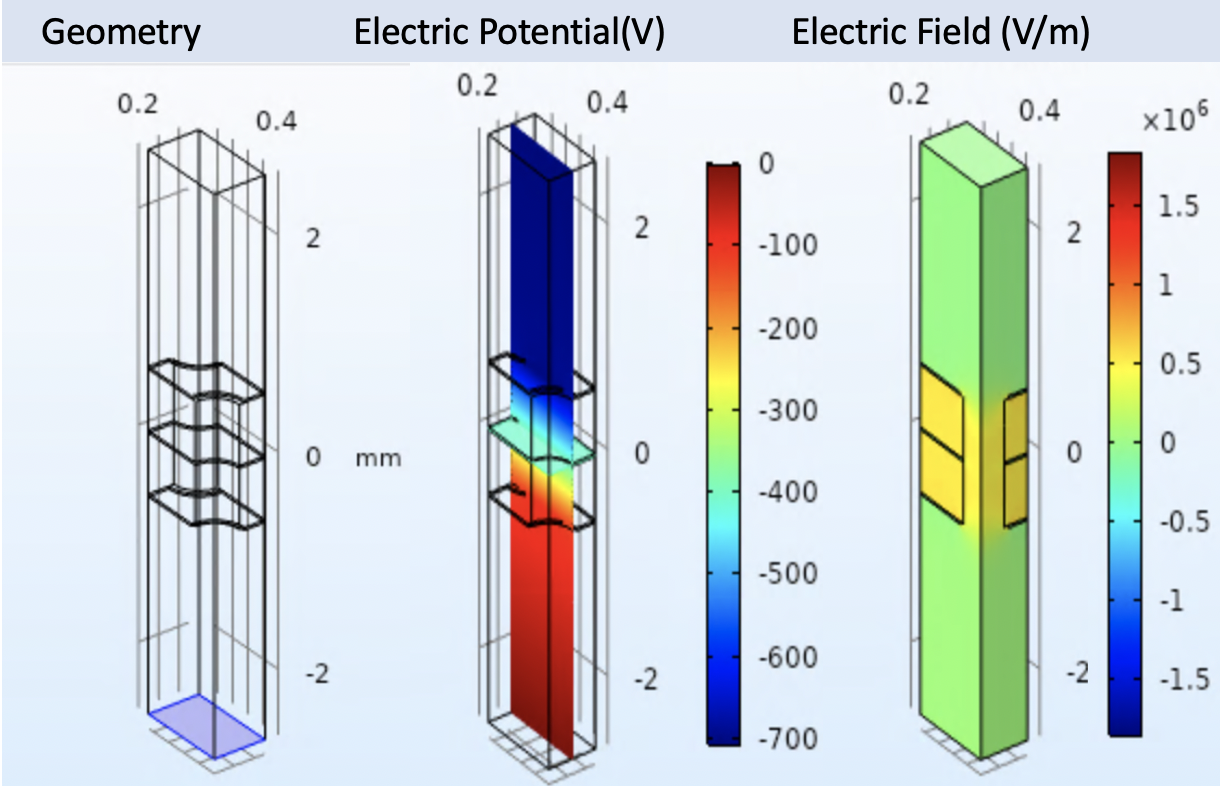}
\caption{Simulated THGEM geometry (left) and voltages (center) as well as electric fields (right) using COMSOL.}
\label{figure5}
\end{figure}
\begin{figure}
\centering
\includegraphics[width=\linewidth]{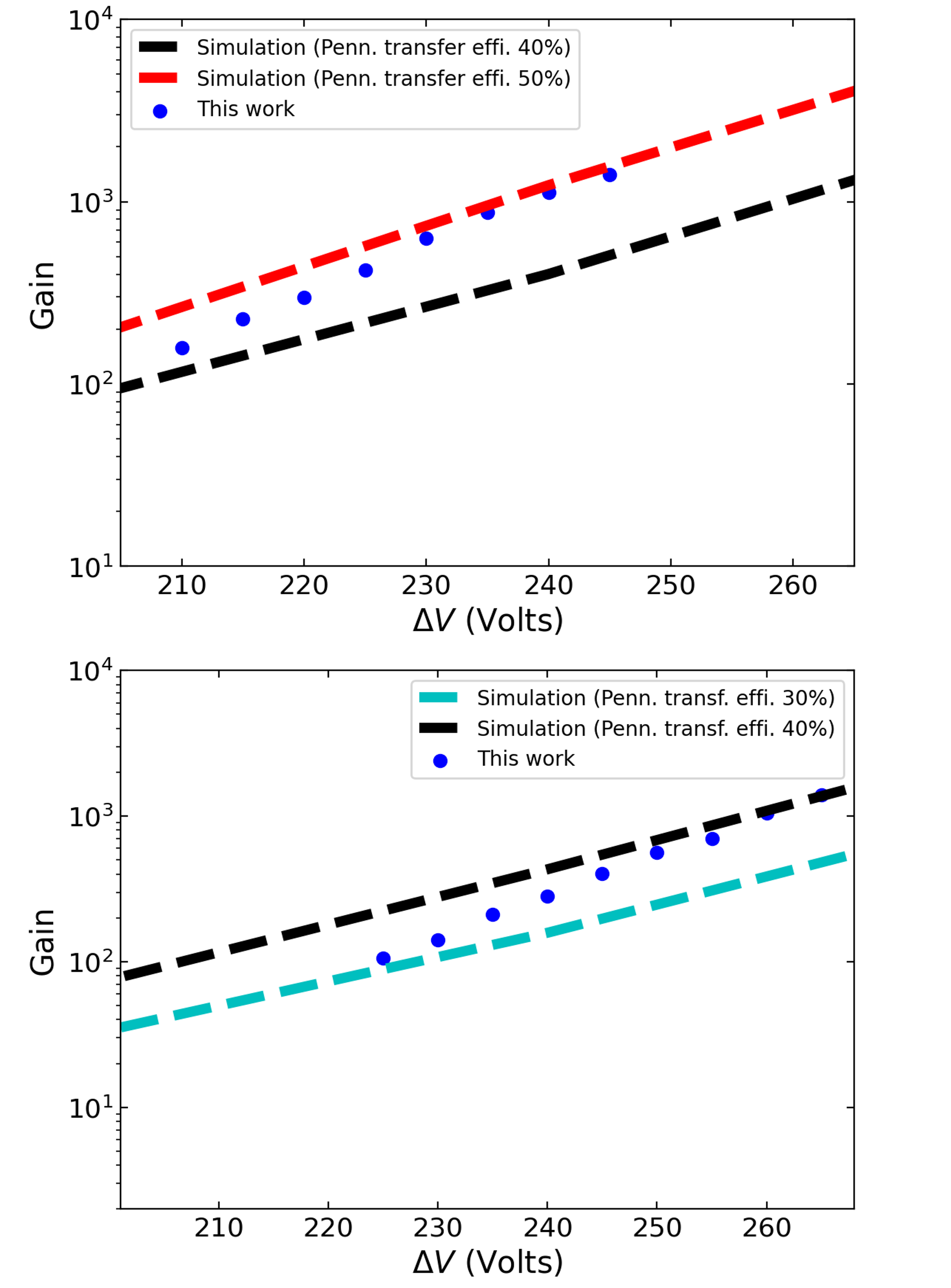}
\caption{Comparison of measured gain with simulated gain at Ne:H$_{2}$ (98:2) mixture (upper panel) and Ne:H$_{2}$ (95:5) mixture (lower panel) at 200 Torr.}
\label{figure6}
\end{figure}

 To understand the amplification and transport of electrons for various THGEM settings, we have performed detailed simulations using Garfield++ \cite{garf}. The whole detector can be divided into three major regions: the drift region between the cathode and top electrode of the THGEM, the amplification region in the THGEM, and the transfer region between the bottom THGEM electrode and the pad plane. In our simulations, a region 2 mm above the top electrode of THGEM and 2 mm below the bottom THGEM holes was considered. The THGEM geometry and electric fields in the drift, amplification, and transfer regions were simulated using the finite-element analysis program COMSOL \cite{Comsol} with the boundary conditions for the potentials set as those used for our measurement. Figure ~\ref{figure5} shows the geometry we used and a simulated electric potential and electric field for one of our simulations. The electric field in the drift region used in the measurements and in the simulation was 30 V/cm and the field in the amplification region varied from 3750 V/cm to 5333 V/cm. The simulated electric fields were imported to Garfield++ \cite{garf} where electron drift and amplification were simulated. We used Garfield++ to calculate the induced charge on the top electrode of the THGEM, which we integrated to compare to the signals we obtained from our measurements. All the parameters in the simulations were fixed by our measurements except for one free parameter, the Penning transfer efficiency, which is the probability of transferring the excitation energy of a metastable state of Ne into a H$_{2}$ ionization in a collision. The Penning transfer efficiency was varied from 30\% to 50\%  to match the gains resulting from the simulations of our data. Figure~\ref{figure6} shows the comparison of simulated gain and measured gain for the Ne:H$_{2}$ (98:2) gas mixture (upper panel) and for the Ne:H$_{2}$ (95:5) gas mixture (lower panel), both at 200 Torr. The comparison shows that the measured gain is best matched by the simulations if the Penning transfer efficiency is between 40\%-50\% for the Ne:H$_{2}$ (98:2), whereas it is between 30\%-40\% for the Ne:H$_{2}$ (95:5). In Garfield++, the Penning effect is simulated by effectively increasing the Townsend coefficient. Therefore an increase in the Penning transfer efficiency increases the Townsend coefficient based on the efficiency value that is used. The lower values of the Penning transfer efficiency were needed for higher concentrations of H$_2$ to match the data. The data show the overall gain is reduced as one increases the H$_{2}$ content in the gas from 2\% to 5\%. This is decrease is reflected in the measured reduced Townsend coefficients shown in Section~\ref{subsec3_1}. As seen in Figure~\ref{figure6}, the slope of the measured gain is higher compared to the simulated gain. This over-exponential growth has been observed in previous GEM-based studies in Penning mixtures and has been attributed to secondary avalanches initiated outside the multiplication region \cite{_ahin_2017}.

\section{Summary and Outlook}
\label{sec5}
There is a particular interest in the low-energy nuclear-physics community in low-pressure operations of pure Ne or Ne-based gas mixtures, especially those mixtures which can minimize background fusion reactions. Ne+H$_{2}$ forms a Penning mixture since the ionization potential of H$_{2}$ is lower than the energy of the metastable state in Ne, creating ionization electrons from the H$_2$ in a fraction of its collisions with Ne. Therefore, it is expected to provide a higher gain compared to pure Ne gas. \color{black} One major advantage of such a gas mixture is that H$_{2}$ in this mixture leads to minimal background reactions compared to other Ne-based gas mixtures such as Ne+CH$_{4}$ and Ne+iC$_{4}$H$_{10}$, where the carbon in these molecules leads to background fusion reactions. Our gain measurements show that gains of more than 10$^{3}$ are achieved at low pressures, i.e. at 150, 200 and 300 Torr in a Ne:H$_{2}$ (95:5) gas mixture and at 200 and 300 Torr for a Ne:H$_{2}$ mixture. Reduced Townsend coefficients for the Ne+H$_{2}$ (95:5) and (98:2) mixtures  obtained from measured gains are enhanced compared to pure Ne. Moreover, they are higher for a 2\% mixture compared with a 5\% mixture, consistent with previously measured Townsend coefficients, in particular the work of Chanin et al.~\cite{Chanin1964}. To understand the electron amplification, transport, and role of the Penning transfer efficiency for the THGEM,  electric fields for the THGEM geometry were simulated using COMSOL, and the electron transport and amplification were simulated using Garfield++. The values of the measured gains were reproduced qualitatively by the simulations. Different values of the Penning transfer efficiency were required for the different H$_{2}$ concentrations to reproduce the observed gains that show a lower H$_{2}$ concentration results in higher gains. This indicates that there is more microscopic physics that is not captured in the simulations, but qualitative descriptions can be obtained for reasonable values of the Penning transfer efficiency. In addition, to demonstrate the use of these gas mixtures in active-target detectors, we imaged 3.2 MeV $\alpha$-particles from $^{148}$Gd in the ND-Cube detector, to demonstrate that the gains achieved are sufficient for measuring  $alpha$-particle tracks on anode pads with an area of 0.4 cm$^2$. Successful imaging of $\alpha$-particles together with the enhanced gain compared to pure Ne gas as inferred from the reduced Townsend coefficient comparison, demonstrates the effectiveness of using Ne+H$_{2}$ gas mixtures as an alternative to pure Ne gas.


Our current study shows that Garfield++ simulations can qualitatively reproduce the measured gain in Ne:H$_{2}$ mixtures. A reasonable quantitative agreement can be achieved with an appropriate choice of the Penning transfer efficiency. However, the  simple modeling of the Penning effect in Garfield++ is not sufficient to explain the observation that lower H$_2$ concentrations result in a higher overall gain. It would be interesting to investigate the possible physical mechanisms for this observation and look for ways to model them.
Another aspect that warrants further investigation is the observed slope of our measured gains, which is larger than what Garfield++ simulations predict and whether it can be accurately modeled by additional known processes.
Last, the current study uses a two-layer M-THGEM, but it would be interesting to see if measurements with a triple-layer M-THGEM  in such a mixture would yield the same features of the gain curves we observe.
\color{black}

\section*{Acknowledgments}
\textit{This work was supported by the National Science Foundation Grant No. 2011890.}

\bibliography{mybibfile}

\begin{thebibliography}{10}
\expandafter\ifx\csname url\endcsname\relax
  \def\url#1{\texttt{#1}}\fi
\expandafter\ifx\csname urlprefix\endcsname\relax\def\urlprefix{URL }\fi
\expandafter\ifx\csname href\endcsname\relax
  \def\href#1#2{#2} \def\path#1{#1}\fi

\bibitem{Bazin}
D.~Bazin, T.~Ahn, Y.~Ayyad, S.~Beceiro-Novo, A.~Macchiavelli, W.~Mittig, J.~S.
  Randhawa,
  \href{https://www.sciencedirect.com/science/article/pii/S0146641020300375}{Low
  energy nuclear physics with active targets and time projection chambers},
  Progress in Particle and Nuclear Physics 114 (2020) 103790.
\newblock \href {http://dx.doi.org/https://doi.org/10.1016/j.ppnp.2020.103790}
  {\path{doi:https://doi.org/10.1016/j.ppnp.2020.103790}}.
\newline\urlprefix\url{https://www.sciencedirect.com/science/article/pii/S0146641020300375}

\bibitem{Ayyad2018}
Y.~Ayyad, D.~Bazin, S.~Beceiro-Novo, M.~Cortesi, W.~Mittig,
  \href{https://doi.org/10.1140/epja/i2018-12557-7}{Physics and technology of
  time projection chambers as active targets}, The European Physical Journal A
  54~(10) (2018) 181.
\newblock \href {http://dx.doi.org/10.1140/epja/i2018-12557-7}
  {\path{doi:10.1140/epja/i2018-12557-7}}.
\newline\urlprefix\url{https://doi.org/10.1140/epja/i2018-12557-7}

\bibitem{Sauli97}
F.~Sauli, {GEM}: A new concept for electron amplification in gas detectors,
  Nuclear Instruments and Methods in Physics Research Section A: Accelerators,
  Spectrometers, Detectors and Associated Equipment 386~(2) (1997) 531 -- 534.

\bibitem{Breskin2009}
A.~Breskin, R.~Alon, M.~Cortesi, R.~Chechik, J.~Miyamoto, V.~Dangendorf,
  J.~Maia, J.~D. Santos,
  \href{http://www.sciencedirect.com/science/article/pii/S0168900208012047}{A
  concise review on thgem detectors}, Nuclear Instruments and Methods in
  Physics Research Section A: Accelerators, Spectrometers, Detectors and
  Associated Equipment 598~(1) (2009) 107 -- 111, instrumentation for Collding
  Beam Physics.
\newblock \href {http://dx.doi.org/https://doi.org/10.1016/j.nima.2008.08.062}
  {\path{doi:https://doi.org/10.1016/j.nima.2008.08.062}}.
\newline\urlprefix\url{http://www.sciencedirect.com/science/article/pii/S0168900208012047}

\bibitem{Cortesi2007}
M.~Cortesi, R.~Alon, R.~Chechik, A.~Breskin, D.~Vartsky, V.~Dangendorf,
  \href{https://doi.org/10.1088%2F1748-0221%2F2%2F09%2Fp09002}{Investigations
  of a {THGEM}-based imaging detector}, Journal of Instrumentation 2~(09)
  (2007) P09002--P09002.
\newblock \href {http://dx.doi.org/10.1088/1748-0221/2/09/p09002}
  {\path{doi:10.1088/1748-0221/2/09/p09002}}.
\newline\urlprefix\url{https://doi.org/10.1088%2F1748-0221%2F2%2F09%2Fp09002}

\bibitem{White05}
J.~T. {White}, J.~{Gao}, J.~{Maxin}, J.~{Miller}, G.~{Salinas}, H.~{Wang},
  {SIGN, a WIMP detector based on high pressure gaseous neon}, in: H.~V.
  {Klapdor-Kleingrothaus}, R.~{Arnowitt} (Eds.), Dark matter in astro- and
  particle physics, 2005, pp. 276--284.

\bibitem{Galea07}
R.~Galea, J.~Dodd, W.~Willis, P.~Rehak, V.~Tcherniatine, Light yield
  measurements of gem avalanches at cryogenic temperatures and high densities
  in neon based gas mixtures., in: 2007 IEEE Nuclear Science Symposium
  Conference Record, Vol.~1, 2007, pp. 239--241.
\newblock \href {http://dx.doi.org/10.1109/NSSMIC.2007.4436322}
  {\path{doi:10.1109/NSSMIC.2007.4436322}}.

\bibitem{Buz2005}
A.~Buzulutskov, J.~Dodd, R.~Galea, Y.~Ju, M.~Leltchouk, P.~Rehak,
  V.~Tcherniatine, W.~Willis, A.~Bondar, D.~Pavlyuchenko, R.~Snopkov,
  Y.~Tikhonov,
  \href{https://www.sciencedirect.com/science/article/pii/S0168900205010831}{Gem
  operation in helium and neon at low temperatures}, Nuclear Instruments and
  Methods in Physics Research Section A: Accelerators, Spectrometers, Detectors
  and Associated Equipment 548~(3) (2005) 487--498.
\newblock \href {http://dx.doi.org/https://doi.org/10.1016/j.nima.2005.04.066}
  {\path{doi:https://doi.org/10.1016/j.nima.2005.04.066}}.
\newline\urlprefix\url{https://www.sciencedirect.com/science/article/pii/S0168900205010831}

\bibitem{Avila16}
M.~L. Avila, K.~E. Rehm, S.~Almaraz-Calderon, P.~F.~F. Carnelli, B.~DiGiovine,
  H.~Esbensen, C.~R. Hoffman, C.~L. Jiang, B.~P. Kay, J.~Lai, O.~Nusair, R.~C.
  Pardo, D.~Santiago-Gonzalez, R.~Talwar, C.~Ugalde,
  \href{https://doi.org/10.1051/epjconf/201611708009}{Study of the
  $^{20,22}${Ne}+$^{20,22}${Ne} and $^{10,12,13,14,15}${C}+$^{12}${C} fusion
  reactions with music}, EPJ Web of Conferences 117 (2016) 08009.
\newblock \href {http://dx.doi.org/10.1051/epjconf/201611708009}
  {\path{doi:10.1051/epjconf/201611708009}}.
\newline\urlprefix\url{https://doi.org/10.1051/epjconf/201611708009}

\bibitem{Yakovlev10}
D.~G. Yakovlev, M.~Beard, L.~R. Gasques, M.~Wiescher,
  \href{https://link.aps.org/doi/10.1103/PhysRevC.82.044609}{Simple analytic
  model for astrophysical $s$ factors}, Phys. Rev. C 82 (2010) 044609.
\newblock \href {http://dx.doi.org/10.1103/PhysRevC.82.044609}
  {\path{doi:10.1103/PhysRevC.82.044609}}.
\newline\urlprefix\url{https://link.aps.org/doi/10.1103/PhysRevC.82.044609}

\bibitem{Beard10}
M.~Beard, A.~Afanasjev, L.~Chamon, L.~Gasques, M.~Wiescher, D.~Yakovlev,
  \href{https://www.sciencedirect.com/science/article/pii/S0092640X10000331}{Astrophysical
  s factors for fusion reactions involving {C}, {O}, {Ne}, and {Mg} isotopes},
  Atomic Data and Nuclear Data Tables 96~(5) (2010) 541--566.
\newblock \href {http://dx.doi.org/https://doi.org/10.1016/j.adt.2010.02.005}
  {\path{doi:https://doi.org/10.1016/j.adt.2010.02.005}}.
\newline\urlprefix\url{https://www.sciencedirect.com/science/article/pii/S0092640X10000331}

\bibitem{Kolata}
J.~Kolata, A.~Howard, W.~Mittig, T.~Ahn, D.~Bazin, F.~Becchetti,
  S.~Beceiro-Novo, Z.~Chajecki, M.~Febbrarro, A.~Fritsch, W.~Lynch, A.~Roberts,
  A.~Shore, R.~Torres-Isea,
  \href{https://www.sciencedirect.com/science/article/pii/S0168900216304181}{Fusion
  studies with low-intensity radioactive ion beams using an active-target time
  projection chamber}, Nuclear Instruments and Methods in Physics Research
  Section A: Accelerators, Spectrometers, Detectors and Associated Equipment
  830 (2016) 82--87.
\newblock \href {http://dx.doi.org/https://doi.org/10.1016/j.nima.2016.05.036}
  {\path{doi:https://doi.org/10.1016/j.nima.2016.05.036}}.
\newline\urlprefix\url{https://www.sciencedirect.com/science/article/pii/S0168900216304181}

\bibitem{Cortesi09}
M.~Cortesi, V.~Peskov, G.~Bartesaghi, J.~Miyamoto, S.~Cohen, R.~Chechik, J.~M.
  Maia, J.~M.~F. dos Santos, G.~Gambarini, V.~Dangendorf, A.~Breskin,
  \href{https://doi.org/10.1088/1748-0221/4/08/p08001}{{THGEM} operation in
  {Ne} and {Ne}/{CH}4}, Journal of Instrumentation 4~(08) (2009)
  P08001--P08001.
\newblock \href {http://dx.doi.org/10.1088/1748-0221/4/08/p08001}
  {\path{doi:10.1088/1748-0221/4/08/p08001}}.
\newline\urlprefix\url{https://doi.org/10.1088/1748-0221/4/08/p08001}

\bibitem{Ne_iso}
F.~Iguaz, S.~Andriamonje, F.~Belloni, E.~Berthoumieux, M.~Calviani, T.~Dafni,
  D.~Oliveira, E.~Ferrer-Ribas, J.~Galáan, J.~Garcáıa, I.~Giomataris,
  C.~Guerrero, Gunsing, D.~Herrera, I.~Irastorza, T.~Papaevangelou,
  A.~Rodráıguez, A.~Tomáas,
  \href{https://www.sciencedirect.com/science/article/pii/S1875389212017233}{New
  developments in micromegas microbulk detectors}, Physics Procedia 37 (2012)
  448--455, proceedings of the 2nd International Conference on Technology and
  Instrumentation in Particle Physics (TIPP 2011).
\newblock \href {http://dx.doi.org/https://doi.org/10.1016/j.phpro.2012.02.392}
  {\path{doi:https://doi.org/10.1016/j.phpro.2012.02.392}}.
\newline\urlprefix\url{https://www.sciencedirect.com/science/article/pii/S1875389212017233}

\bibitem{Cortesi2017}
M.~Cortesi, S.~Rost, W.~Mittig, Y.~Ayyad-Limonge, D.~Bazin, J.~Yurkon,
  A.~Stolz, \href{https://doi.org/10.1063/1.4974333}{Multi-layer thick gas
  electron multiplier (m-thgem): A new mpgd structure for high-gain operation
  at low-pressure}, Review of Scientific Instruments 88~(1) (2017) 013303.
\newblock \href {http://arxiv.org/abs/https://doi.org/10.1063/1.4974333}
  {\path{arXiv:https://doi.org/10.1063/1.4974333}}, \href
  {http://dx.doi.org/10.1063/1.4974333} {\path{doi:10.1063/1.4974333}}.
\newline\urlprefix\url{https://doi.org/10.1063/1.4974333}

\bibitem{Ahn2022}
T.~Ahn, J.~Randhawa, S.~Aguilar, D.~Blankstein, L.~Delgado, N.~Dixneuf,
  S.~Henderson, W.~Jackson, L.~Jensen, S.~Jin, J.~Koci, J.~Kolata, J.~Lai,
  J.~Levano, X.~Li, A.~Mubarak, P.~O’Malley, S.~{Ramirez Martin}, M.~Renaud,
  M.~Serikow, A.~Tollefson, J.~Wilson, L.~Yan,
  \href{https://www.sciencedirect.com/science/article/pii/S0168900221010536}{The
  notre-dame cube: An active-target time-projection chamber for radioactive
  beam experiments and detector development}, Nuclear Instruments and Methods
  in Physics Research Section A: Accelerators, Spectrometers, Detectors and
  Associated Equipment 1025 (2022) 166180.
\newblock \href {http://dx.doi.org/https://doi.org/10.1016/j.nima.2021.166180}
  {\path{doi:https://doi.org/10.1016/j.nima.2021.166180}}.
\newline\urlprefix\url{https://www.sciencedirect.com/science/article/pii/S0168900221010536}

\bibitem{garf}
R.~Veenhof, {Garfield, a drift chamber simulation program}, Conf. Proc.
  C9306149 (1993) 66--71, [,66(1993)].

\bibitem{Pollaco2018}
E.~Pollacco, G.~Grinyer, F.~Abu-Nimeh, T.~Ahn, S.~Anvar, A.~Arokiaraj,
  Y.~Ayyad, H.~Baba, M.~Babo, P.~Baron, D.~Bazin, S.~Beceiro-Novo,
  C.~Belkhiria, M.~Blaizot, B.~Blank, J.~Bradt, G.~Cardella, L.~Carpenter,
  S.~Ceruti, E.~{De Filippo}, E.~Delagnes, S.~{De Luca}, H.~{De Witte},
  F.~Druillole, B.~Duclos, F.~Favela, A.~Fritsch, J.~Giovinazzo, C.~Gueye,
  T.~Isobe, P.~Hellmuth, C.~Huss, B.~Lachacinski, A.~Laffoley, G.~Lebertre,
  L.~Legeard, W.~Lynch, T.~Marchi, L.~Martina, C.~Maugeais, W.~Mittig,
  L.~Nalpas, E.~Pagano, J.~Pancin, O.~Poleshchuk, J.~Pedroza, J.~Pibernat,
  S.~Primault, R.~Raabe, B.~Raine, A.~Rebii, M.~Renaud, T.~Roger,
  P.~Roussel-Chomaz, P.~Russotto, G.~Saccà, F.~Saillant, P.~Sizun, D.~Suzuki,
  J.~Swartz, A.~Tizon, A.~Trifiró, N.~Usher, G.~Wittwer, J.~Yang,
  \href{https://www.sciencedirect.com/science/article/pii/S0168900218300342}{Get:
  A generic electronics system for tpcs and nuclear physics instrumentation},
  Nuclear Instruments and Methods in Physics Research Section A: Accelerators,
  Spectrometers, Detectors and Associated Equipment 887 (2018) 81--93.
\newblock \href {http://dx.doi.org/https://doi.org/10.1016/j.nima.2018.01.020}
  {\path{doi:https://doi.org/10.1016/j.nima.2018.01.020}}.
\newline\urlprefix\url{https://www.sciencedirect.com/science/article/pii/S0168900218300342}

\bibitem{Chanin1964}
L.~M. Chanin, G.~D. Rork, Primary ionization coefficient measurements in
  penning mixtures, Physical Review 135 (1964) 71--75.

\bibitem{Kruithof1937}
A.~Kruithof, F.~Penning,
  \href{https://www.sciencedirect.com/science/article/pii/S0031891437800750}{Determination
  of the townsend. ionization coefficient $\alpha$ for mixtures of neon and
  argon}, Physica 4~(6) (1937) 430--449.
\newblock \href
  {http://dx.doi.org/https://doi.org/10.1016/S0031-8914(37)80075-0}
  {\path{doi:https://doi.org/10.1016/S0031-8914(37)80075-0}}.
\newline\urlprefix\url{https://www.sciencedirect.com/science/article/pii/S0031891437800750}

\bibitem{Comsol}
C.~Multiphysics, Introduction to comsol multiphysics{\textregistered}, COMSOL
  Multiphysics, Burlington, MA, accessed Feb 9 (1998) 2018.

\bibitem{_ahin_2017}
O.~Sahin, T.~Kowalski, \href{https://doi.org/10.1088/1748-0221/12/01/c01035}{A
  comprehensive model of penning energy transfers in ar - {CO}2mixtures},
  Journal of Instrumentation 12~(01) (2017) C01035--C01035.
\newblock \href {http://dx.doi.org/10.1088/1748-0221/12/01/c01035}
  {\path{doi:10.1088/1748-0221/12/01/c01035}}.
\newline\urlprefix\url{https://doi.org/10.1088/1748-0221/12/01/c01035}

\end{thebibliography}
\end{document}